# BLOCK-CHAIN TECHNOLOGIES IN HEALTHCARE ANALYTICS

**Fathima Begum M[1], Subhashini Narayan[2]**
*Assistant Professor Junior [1], Assistant Professor Senior[2]*
*School of Information Technology and Engineering,*
*Vellore Institute of Technology*

*ABSTRACT*

*Research has advanced to broaden its applications to cases of non-financial usage after the block-chain was presented by Bitcoin. Healthcare is one of the sectors in which block-chain has tremendous impacts. Exploration here is generally new yet developing quickly; along these lines, health informatics researchers and specialists are continually battling to stay up with research progress around there. The block-chain's information accessibility, protection and security characteristics have an auspicious future in healthcare, providing solutions to the healthcare framework's multifaceted nature, classification, integrity, interoperability and protection issues. This paper focuses on a scoping analysis of the current research into the use of healthcare system block-chain innovation.*

## 1. INTRODUCTION

The use of paper-based clinical records has routinely shaped conventional healthcare services rehearsals and became Electronic patient records. As a result, Electronic Health Records (EHRs) also contain extremely sensitive clinical information, which is split between providers of medical care, drug stores and patients. Present EHR approaches by managers include disseminated or cloud server data storage, which may cause various features and monetary problems and are often helpless against various cyber security threats, including malware and ransom ware [1].

Perhaps the greatest riddle of humankind is the conservation and development of information. People try to give collected innovative and logical accomplishments for people in the future through permanent records. In old occasions, civic establishments utilized oral conventions, cavern artistic creations and stone carvings before they chose papyrus leaves. Papyrus is currently more broadly known [2]. It has been recommended that computerized recording of circles, documents and the Internet would be a superior arrangement, however up until this point, the chronicle is inadequate: disk failure, file system failure and website misfortune make advanced capacity an uncommon work . Presently, we have another innovation that is required to tackle every one of these issues, called Block-chain. Be that as it may, people would especially prefer not to record information for their own advantage, however record information, so information can be extended as a transformative endurance system [3].

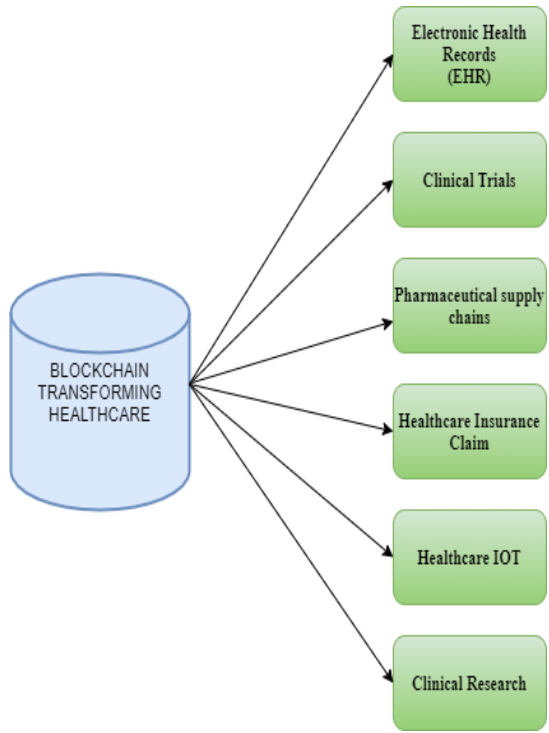

Figure 1: Block-chain in healthcare applications

The advancement of computerized innovation has changed the amount and nature of clinical information from hereditary data to clinical records [4]. These progressions require improved possession and security, protection, honesty, and detect ability in the creation, stockpiling, transmission, sharing, and usage of clinical information. Given the anticipated progress of the IoT in the coming years, it is important to place trust in this immense source of knowledge that is approaching. As a key innovation that will alter the way we share data, Block-chain has emerged. Building trust in suitable circumstances without specialist requirements is a creative development that can potentially change a range of businesses, including the IoT [5]. For example, big data and distributed computing have been used by IoT to solve its constraints since its inception, and we think block-chain would be one of the next ones.

The rest of the paper is organized as follows: Section 2 explains Literature review; Section 3 introduces the Block-chain technology and analyzes its working. In Section 4 various Consensus algorithms are explained and the application of Block-chain in healthcare systems and its challenges are described in section 5. Lastly, our conclusion is presented in Section 6.

## 2. LITERATURE REVIEW

Iyolita et al., [6] focused on planning and creating calculated structures to assemble more solid, straightforward and effective digital systems. While block-chain brings a wide assortment of advantages, it additionally forces certain challenges. Also, to comprehend the properties of block-chain, its present uses, noticed advantages and entanglements to give a reasonable comprehension of Block-chain.

Tariq et al., [7] explained IoT-based smart healthcare systems have enormously increased the value of the medical care area with the utilization of wearable and cell phones. This prompts a significant utilization of health data sharing for the improved, precise, and opportune diagnosis. In any case, smart healthcare systems are profoundly powerless against a few security breaks and different threatening assaults, for example, protection spillage, hardening, fabrication etc. It additionally amalgamates the possibilities of block-chain innovation as a promising safety effort, features likely difficulties in the healthcare domain. Usman et al., [8] proposed a prototype of the Electronic Medical Records Management

System using the approved block-chain network "Hyper ledger". Using the encoded private key and public key, the patient's data will be put away. The private key may simply be decrypted by the biometric signature of the patient. Using a mix of public and private keys, critical patient records can only be accessed in emergency situations.

| References | Year | Summary of concept |
|---|---|---|
| *Iyolita et al.,* | 2020 | Designed and Developed conceptual digital framework using Block-chain system. |
| *Tariq et al.,* | 2020 | IoT-based smart healthcare systems with the utilization of wearable and cell phones |
| *Usman et al.,* | 2020 | For successful monitoring and exchange of electronic medical records, a prototype of the Electronic Medical Records Management System using the approved Block-chain network "Hyper ledger" is used. |
| *Kim et al.,* | 2020 | Established artificial intelligence platform for information security block-chain and validated block-chain systems for accurate data extraction, storage, and verification |
| *Jennath et al.,* | 2020 | In e-Health, where a transparent framework for consent-based data sharing is built, the potential of developing trusted Artificial Intelligence models over block-chain has been explored. |
| *Zghaibeh et al.,* | 2020 | Developed Smart-Health (SHealth), a Block-chain based health management. |
| *Chung et al.,* | 2019 | Proposed knowledge based block-chain network for mobile service management of health log data using the side chain structure |
| *Dagher et al.,* | 2018 | Ancile has introduced a platform that uses smart contracts for improved access control in an Ethereum based block-chain and uses advanced cryptographic techniques for further security. |
| *Zhang et al.,* | 2018 | Proposed a Safe and Privacy-Preserving PHI Sharing (BSPP) block-chain based framework for diagnostic improvements in e-health systems |
| *Li et al.,* | 2018 | Developed a novel medical block-chain-based data preservation system (DPS). |

Table 1. Summary of Literature review

Kim et al., [9] described an artificial intelligence platform for information security block-chain and validated block-chain systems for accurate data extraction, storage, and verification. Furthermore, to obtain the TPS of medical data, different verification and performance assessment metrics are set. It maximizes block-chain secrecy and the sensitivity and usability of artificial intelligence. A block-chain based solution for health records was implemented by Jennath et al., [10] to address the protection and privacy problems that are currently not present in existing e-Health systems. In e-Health over block-chain, where a transparent network is developed for the sharing of consent-based data, this work also explores the potential of creating trustworthy Artificial Intelligence models.

Smart-Health (SHealth) was developed by Zghaibeh et al., [11] it is an integrated multi-layered private block-chain with a multi-layered addressing mechanism that defines the rights and permissions of the entities of the system. Block-chain-based systems guarantee anonymity, transparency, availability, tampering and malicious attack resistance, smooth integration and easy management of data. Chung et al., [12] recommended the mobile health log information management service's information-based block chain network. The log data and setting data of the customer are applied to block-chain technology that is difficult to generate and distort in the information-based health platform, enabling many customers to record information and set data collected continuously to be placed in a block in the information base using the side chain structure that stores data through the setup of knowledge-based information exchange.

Dagher et al., [13] introduced a system called Ancile that uses smart contracts for improved data access control and deception in an Ethereum-based block-chain, and uses advanced cryptographic security techniques, as well as exploring how Ancile can interact with the various needs of patients, providers and third parties, and recognizing how the framework could solve long-standing privacy. A Block-chain-based safe privacy preservation sharing framework for electronic clinical records was proposed by Zhang et al., [1]. This strategy uses two Block-chains: the private Block-chain and the Block-chain consortium. The private block-chain holds patient's health data, and the block-chain consortium makes protected files for the details that the private block-chain puts away. Both general health records and the identity of patients are encrypted by public key with keyword search to make the plan information safe.

Li et al., [14] faced an earnest need to digitize the records in the medical care industry on top of security and confidentiality of patient's details. To fuse development in the field of the medical services industry, they proposed a system to manage the medical services knowledge using the block-chain innovation. In their work, they accomplish validation, classification, and responsibility for the necessary information sharing. The proposed framework accomplishes the decentralization of clinical records in a made sure about way. Their structure utilized the brilliant agreement instrument and POW–based agreement calculation to approve another block in their block-chain-based framework.

## 3. BLOCK-CHAIN BACKGROUND STUDY

Block-chain arose about 10 year's prior and gathered consideration as the "distributed ledger" technology that goes about as the spine to the digital currency bit coin. Medical care has woken up to Block-chain's abilities in the course of recent years as the innovation has advanced past bit coin to address the issues of profoundly controlled organizations. The United States spends about 20% of its GDP on medical care. Block-chain is an innovation that is now getting enormous consideration in medical care. In addition, as per a study by BIS Research, global healthcare industry spending on Block-chain is required to reach $5.61 billion by 2025[17].

### 3.1 Block-chain Structure

These are the center Block-chain components.

Node - device or PC within the architecture of the Block-chain (each has an independent copy of the entire Block-chain record). Transaction - the smallest building block of a Block-chain system that fills in as the reason for the Block-chain (records, data, and so on). Block - an information system used to establish a collection of exchanges that are suitable for all organizational nodes. Chain - in a specific request, a succession of blocks. Miners - clear nodes that play out the measure of block validation before contributing something to the structure of the Block-chain. Consensus (Agreement Convention) - a bunch of guidelines for completing Block-chain tasks and plans. The structure of another block is indicated by every new record or exchange within the Block-chain. To guarantee its authenticity, each record is then demonstrated and carefully endorsed. This block should be tested by most nodes in the system before it is connected to the network [18].

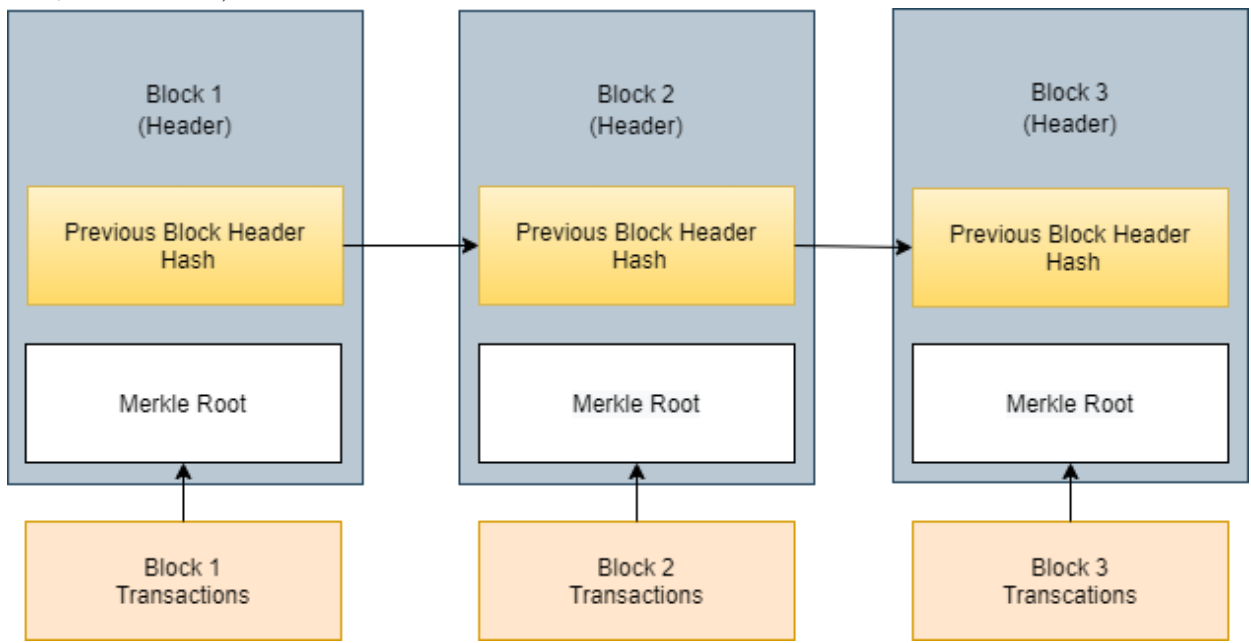

Figure2: Block-chain Structure

Either permission-less or permission-less can be delegated to a Block-chain network. Using a pseudo-name, the permission-less network (called as public) is available for everyone to join. Clients are urged to join the network by providing impetus for enhanced security. The record data is visible to everyone in this organization. A greeting-based organization that allows only authorized support is the permitted network (otherwise called private). As characterized by the endorsement authority, the perceivability of the information is responsible for acquiring control rights. The authorized Block-chain is high flexible and has a greater throughput compared to the authorization less, speaking about the restricted organizational cooperation.

## 4. CONSENSUS ALGORITHMS

The idea driving Block-chain is a made sure about and confided in engineering because of network consensus [19]. Distinctive consensus algorithms have been actualized in the past for explicit applications on the grounds that every area has explicit prerequisites. For instance, a few spaces require low calculation power, while others require quicker preparing of exchange. The vital capacity of Block-chain innovation is consensus algorithms, which delineates the calculation expected to agree between network hubs during blocks confirmation measure. Consensus algorithms plan to give correspondence between excavators, giving similar load to every one of them with the goal that larger part of the diggers can arrive at a choice [20].

To uphold the accuracy and reliability of knowledge, Block-chain has adopted decentralized consensus algorithms. Examples of consensus algorithms include

### 4.1 Proof of Work (PoW)

Proof of Work (PoW) follows an idea of work, where it depends on the way that nodes are less inclined to assault the network on the off chance that they play out a ton of work. PoW-based Block-chain expects diggers to perform computationally costly errands (completed by numerous substances) to add a block to the Block-chain, in this manner making it practically unimaginable for Sybil assaults [21]. PoW operates in a manner called mining; before a response is sought, nodes will execute figures. For example, the estimation period in the Bitcoin block-chain aims to find an odd number (called nonce) to generate the correct block header hash. Miners must be able to play out a particular measure of work along these lines to decide the amount. All separate nodes are responsible for verifying that at the point where the miner takes care of the problem, the response needed is correct. PoW burns more fuel, making it inefficient to be used in low-force applications.

### 4.2 Proof of Stake (PoS)

Proof of Stake (PoS) divides clients by the block-chain's stake. A miner can be any node that has a particular measure of stake in the block-chain. This consensus algorithm recognizes that there is a lower risk for a customer with more stakes to target the network. When being a miner, nodes distribute a basic measure of their stake [22]. Consequently, to ensure that a client is trusted and allowed to mine, the network will keep the amount up. As it needs less computational force, PoS have lower energy consumption than PoW. The problem with PoS is that the Block-chain mining cycle focuses on the richest members, since they can claim a higher stake than distinct nodes.

### 4.3 Delegated Proof of Stake (DPoS)

Delegated Proof of Stake is another consensus algorithm suggested to upgrade PoS. Those agents are responsible for the scheme in this algorithm, rather than assigning the partners the generation and approval of blocks. Faster exchanges, since fewer nodes are used, are one of the benefits of this consensus algorithm. Furthermore, the selected nodes may change the block size and intervals.

### 4.4 Transactions as Proof of Stake (TaPoS)

Transactions as Proof of Stake are a PoS variation (TaPoS). Unlike PoS, where some nodes contribute to the security of the network, all TaPoS nodes contribute to the security system. In PoS, the impediment is due to the age of the stake that is acquired in any case where the node is not linked to the network. To reward nodes on the Block-chain that depend on their movement and possession, Proof of Action (PoA) is proposed.

### 4.5 Practical Byzantine Fault Tolerance (PBFT)

For non-concurrent situations, PBFT was proposed to tackle the issue of the Byzantine Generals. It agrees that more than 2/3 of all nodes are real, while not only a third is malignant. Through each block age, a pioneer is chosen and the pioneer is responsible for requesting exchanges. The approval of the node should help to add a node, at least 2/3, all things considered. Designated BFT (DBFT) is a variation of BFT which also operates with DPoS, where it is responsible for authorizing and generating blocks for a certain number of nodes.

The advantages and drawbacks of different Block-chain consensus algorithms are summarized in Table 2.

| Algorithm | Advantages | Drawbacks |
|---|---|---|
| Proof of Work (PoW) | *Provides detailed power and control decentralization in the network. *More stable network of networks. | *High power of processing (expensive). * Heavy use of electricity. * It's possible to weaken small networks. |
| Proof of Stake (PoS) | * More efficient resources. *Better incentives with greater stakes. *Provides quicker transaction processing. | *Less network that is decentralized than PoW. * Less security than PoW. |
| Delegated Proof of Stake (DPoS) | *Quicker than PoW and PoS processing. * Better delivery of incentives. * Efficiency of Energy. * Lower costs for hardware. | * More vulnerable to attacks. *Affluent people manage the network. *Due to less decentralization, less resilience. |
| Transactions as Proof of Stake | * More protection than PoS, as all nodes in the network | *Lower speed than DPoS because the inclusion of all nodes. |
| (TaPos) | contribute.<br>* Provides a simplified algorithm for PoS. | * Does not work well when the Block-chain forks are short. |
| Practical Byzantine Fault Tolerance (PBFT) | *Ability without the need for confirmation to make transactions such as in PoW.<br>* Major decrease in electricity consumption. | * Due to a high amount of contact between nodes, it works only in limited consensus group sizes.<br>*Impossible to assert the legitimacy of a message to third parties.<br>*Vulnerable to attacks from Sybil. |

Table 2. Advantages and Drawbacks of different Block-chain consensus algorithms

## 5. BLOCK-CHAIN BASED HEALTHCARE SYSTEM

The essential Block-chain stage for medical care which gives insignificant usefulness to program medical services exchanges. This uses an authorized block-chain network less because of its preferences over authorization. The block-chain based medical services model scheme consists of patients, such like patients, allied health practitioners (specialists, attendants and pharmacy specialists) and executive's tools, such as information on clinical tests and exchanges, such as updating and investigating health records [15]. The developed block-chain based medical healthcare platform overview is stated in Figure 3. It demonstrates that the stage involves a few medical clinics and participants affiliated with the block-chain network, including patients.

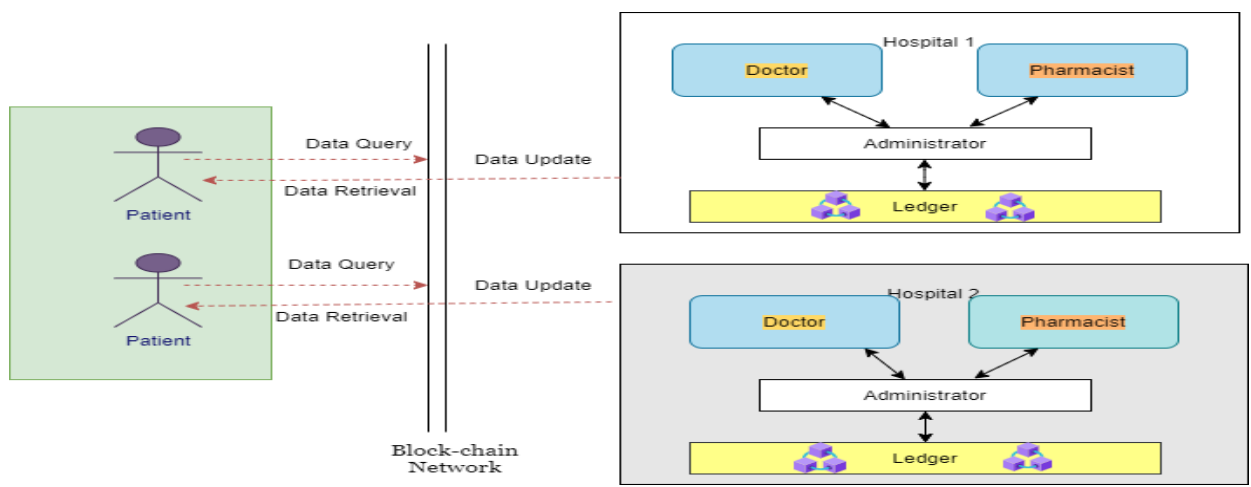

Figure 3: Healthcare Platform Block-chain based.

Let N reflect in the network the total number of hospitals,

X total participants, and

Y total patients (Y < X).

Each hospital i, i€ {1. . . N} retains a replica of the ledger.

A participant k, k€ {1. . . X} updates the queries related to the health record of a patient j, j€ {1 . . . Y}.

The doctors and pharmacist act as complete nodes for each medical care transaction and the administrator works as mining node. The created insignificant Block-chain based medical services stage upholds two sorts of exchanges: (1) information update and (2) information inquiry.

## 5.1 A workflow of block-chain based healthcare applications

Emerging block-chain based healthcare innovations, including data sources, block-chain technology, healthcare applications, and stakeholders, are conceptually organized into four layers [23]. A description of the block-chain based workflow for healthcare applications is illustrated in figure 4.

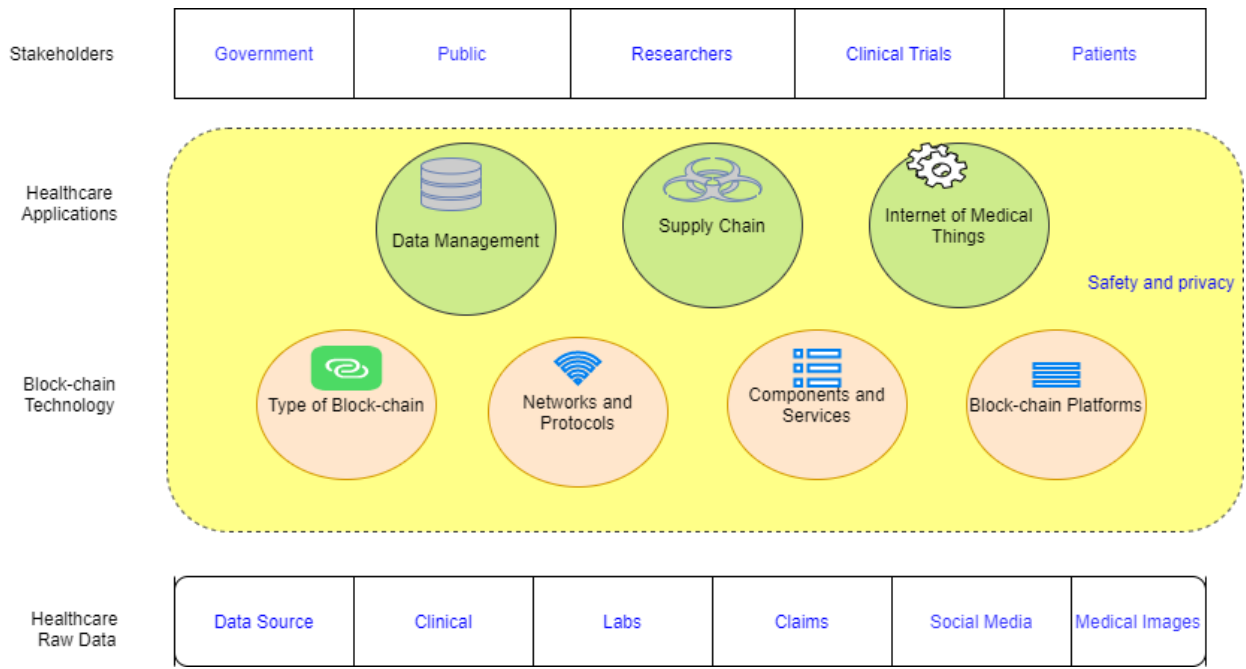

Figure 4: Block-chain based workflow for healthcare applications

Initially, all data is consolidated and raw data is generated from medical equipment's, laboratories, social networks, and many other sources, which subsequently grew to big data in size. The main element of the entire block-chain based healthcare system is this knowledge, and it is the primary element that produces the first layer of the stack. At the top of the raw data layer is the block-chain technology, which is considered as the main mechanism for building a stable healthcare architectural features and further divided into four components. There are various features for each block-chain network, such as consensus algorithms and protocols. Block-chain systems enable the development and management of transactions by users. Several block-chain systems, such as Ethereum, Ripple, and Hyper ledger Fabric were developed and are currently in use. Smart contract, signatures, wallets, activities, digital Assets and membership are the key elements of the block-chain. A broad variety of protocols could be used for communicating with other systems and structures, or even through multiple networks. This can include P2P, centralized, decentralized, and distributed, for instance First, data management, including global research and development (R&D) scientific data exchange, data management, data storage (e.g., cloud-based applications) [24] and EHRs. The second tier, including clinical studies and pharmaceutical drugs, reflects SCM applications. Eventually, the IoMT includes the third tier, namely a combination of IoT and medical devices for healthcare, IoT technology and data protection for healthcare, and AI. Figure 4 in the block-chain reveals applications for healthcare. Finally, at the top of the hierarchy is the stakeholder tier, consisting of groups benefiting from block-chain based healthcare scenarios, including application developers, academics, and patients. At this point, the main concern of users is to efficiently share process and manage information without hindering its security and privacy.

Initially, block-chain technology (BCT) was intended for the most popular execution in the monetary and digital currency fields, however now its utility is extending in a few different fields, including the biomedical field [25]. The potential of block-chain innovation can be seen in various fields like medication, genomics, teletherapy, remote control, e-medication, nephrology and personalization. Through its component of settling and securing informational collections, medical services applications permit clients to interface with them through various kinds of exchanges (as appeared in Figure 4).

These defined block-chain use cases are discussed in more depth in the following subsections.

## 5.2 Block-chain in Pharmaceutical supply chains

One of the most common block-chain based use cases in healthcare is Pharmaceutical supply chains. Block-chain in healthcare services is tied in with eliminating the middleman. Block-chain innovation can genuinely upset the pharmaceutical sector in general, just as drug disclosure, improvement, and appropriation measure [26]. That all by itself isn't unexpected in any way. We assessed in advance that 9 out of 10 new drugs do not pass the clinical preliminary stage, and many more do not reach the (Food and Drug Administration)FDA-approval stage[27]. In a time when people are not widely trusted, the major challenge for pharmaceutical organizations is having valuable patient information. Obviously, Big Pharma is dealing with a lot of data issues. The following are challenges in Pharmaceutical supply chain

### 5.2.1 Divergence in Pharma Data

Probably the most important information problem considered by drug organizations, particularly with regard to drug growth, is the serious uniqueness of data. Recall pharma information is frequently recorded and concealed in storehouses, and is gotten to through a few distinct stages. What's far more terrible is that every division regularly utilizes a different information structure and model. Accordingly, gaining admittance to exceedingly significant information at whatever point required is essential for pharma organizations for enormous information investigation [28]. This is a zone where pharmaceutical companies can collaborate with Artificial Intelligence, Big data analytics, and Block-chain to communicate expertise and accurate results. Fortunately, with block-chain use in the pipeline, about 61 percent of pharmaceutical companies already recognize Artificial Intelligence.

### 5.2.2 Time-consuming and resource-heavy Data Analytics

Taking into account that pharma organizations need to manage so numerous information sources, obviously there's an issue with regards to information social affair, taking care of, and at last analytics. First of all, some vital information may get lost in the noise during the catching stage, watering down bits of knowledge from the analytics. All things considered, clearly the entire information taking care of cycle for pharma organizations need to move. All things considered, overseeing enormous informational collections is now a serious test for both big and small drug organizations [29].

### 5.3 Block-chain in Healthcare Insurance Claim and Data Analytics

We can find some illustrations of using block-chain for health insurance purposes. They recommend safe health insurance record capacity that can assist emergency clinics and insurance agencies with the need for colossal extra space and security systems. This block-chain comprises of nodes speaking to clinics, insurance companies, workers, and record nodes. Coding and billing enforcement, claim transmission, verifying financial obligation and obtaining payment from insurance providers are all part of the claim process. Since some of the fees are completely covered by the patient's private health care plan or are charged by the patient, the whole billing system may be difficult [16]. Claims and billing in the healthcare industry are systematically abused, but they can be resolved or minimized by introducing a transparent framework that benefits all stakeholders. Block-chain will maintain a transparent framework that keeps everybody involved in the process and removes distrust.

### 5.4 Block-chain in Electronic Health Records (EHR)

In the previous decade experts, emergency clinics, and medical care hardware have encouraged a tremendous interest for the digitization of clinical wellbeing records[30], on the grounds that the digitization of this information makes it simple to access and share, and is the reason for better and quicker dynamic. The most well-known use of block-chain innovation in medical care is right now in the field of electronic clinical consideration.

Notwithstanding, electronic clinical records (EHR)have never been made to manage lifetime records between different foundations, while quiet information is dissipated across establishments since day to day environments separate them from one supplier's information. Along these lines, they won't have the option to effortlessly get to the past information. There is an earnest need to handle electronic clinical records in a creative manner to urge patients to take an interest in their current and chronicled clinical information. Numerous scientists have proposed block-chain innovation to keep up electronic clinical records. The model named "MedRec" uses one of kind block-chain advantages to oversee confirmation, secrecy, uprightness, and simple sharing of information. It takes a shot at a decentralized record the board framework and cases to give patients definite, constant verifiable records, and permits simple admittance to their separate medical services data across different suppliers and therapy organizations [31]. EHR as a rule contains profoundly delicate and basic information identified with the patient, which is frequently shared among clinicians, radiologists, medical care suppliers, drug specialists, and scientists for compelling analysis and therapy. During the time spent putting away, communicating, and disseminating profoundly delicate patient data among various substances, the patient's therapy techniques might be influenced, which may genuinely undermine the patient's wellbeing and keep up the patient's most recent clinical history. For patients with ongoing illnesses, (for example, malignant growth and HIV), due to the long history of when treatment, development and restoration, the predominance of such dangers might be higher. Along these lines, keeping up a modern patient history has become a main concern to guarantee powerful treatment.

### 5.5 Block-chain in Clinical Research.

A progression of issues may emerge in clinical preliminaries, including information protection, information sharing, record keeping; patient enrollment, and so forth Cutting edge Internet, block-chain can give plausible answers for these issues. Medical care analysts are utilizing block-chain innovation to tackle these issues [32]. With computerized reasoning (AI) and machine learning, the use of block-chain will before long range the clinical business. The principle focal point of the investigation is to tackle the issue of patient enrollment. The consequences of the examination show that Ethereum exchanges are quicker than Bit coin, so the ends came to propose that Ethereum smart contracts ought to be utilized in clinical preliminaries to build the straightforwardness of the information the board framework. Thusly, the utilization of block-chain to enroll patients is one of the current uses of this innovation in clinical examination.

### 5.6 Block-chain in Healthcare IOT

The consensus mechanisms used for transaction validations in the block-chain based models mentioned above are computation intensive. For the large proportion of IoT applications, these solutions are not feasible, considering the costs involved in block-chain processing [33]. The integration of smart contracts into the block-chain has brought tremendous momentum to block-chain based applications to revolutionize the application scenarios where it is possible to use the block-chain.

The off-chain system is used to store information in IOT applications because it is expensive to store vast quantities of data across the chain. Access to the off-chain database is managed by smart contracts. IOT devices are initially connected to the patient body to define parameters such as heart rate and body temperature, etc. Sensors read these parameters, processes and store data.

The above mentioned Block-chain application in Healthcare is summarized below

| Applications | Summary |
|---|---|
| Pharmaceutical supply chains | Genuine stock is overseen and the supply chain is followed efficiently utilizing this block-chain innovation. This is tended to by utilizing the smart contract which assists with following the movement of drugs from producer to provider, provider to the affiliate, and affiliate to drug stores lastly drug stores to patient. |
| Healthcare Insurance Claim and Data Analytics | Capacity of Health Insurance Industry is multi-faceted. The specialist organization needs to deal with the claims of policy holders. It is discovered that the level of Transparency, Trust while preparing an exchange, claim settlement time, security of information and so forth are significant for approving a claim effectively. |
| Electronic Health Records | From the data generation stage to the data retrieval stage, without manual interference, the integrity of the encrypted EHR on the shared ledger of the activated block-chain can be ensured. |
| Clinical Research | For any knowledge sharing that could occur in clinical research, Block-chain provides a decentralized security system. |
| Healthcare IOT | IoT empowers medical care professionals to be more attentive and interface with the patients proactively. IoT gadgets labeled with sensors are utilized for following constant area of clinical hardware like wheelchairs, defibrillators, nebulizers, oxygen siphons and other monitoring equipment. |

Table 3. Applications of Block-chain Technology in healthcare use cases.

# 6. CONCLUSION

The aim of the current study was to understand the scope of block-chain application in the healthcare domain. We provided an overview of block-chain and decentralized storage on how block-chain technology can improve the healthcare system. Due to the sensitive nature of data being processed and managed, block-chain technology presents a decentralized network and is considered to have great potential for use in healthcare. The purpose of the study was to identify the current status of healthcare block-chain research and application. Our results suggest that research into block-chain technology and its employment in healthcare is growing. Current trends in healthcare block-chain research indicate that it is mostly used for sharing data, health records and control of access, supply chain management, etc. A novel framework, architecture or model using block-chain technology in healthcare is presented in most research. As for further research, in the field of healthcare, block-chains are still fairly new technologies and new ways of using them can still be found and researched.

[30] Vora, J., Nayyar, A., Tanwar, S., Tyagi, S., Kumar, N., Obaidat, M. S., & Rodrigues, J. J. (2018, December). BHEEM: A block chain-based framework for securing electronic health records. In *2018 IEEE Globecom Workshops (GC Wkshps)* (pp. 1-6). IEEE.

[31] Azaria, A., Ekblaw, A., Vieira, T., & Lippman, A. (2016, August). Medrec: Using block chain for medical data access and permission management. In *2016 2nd International Conference on Open and Big Data (OBD)* (pp. 25-30). IEEE.

[32] Khan, F. A., Asif, M., Ahmad, A., Alharbi, M., & Aljuaid, H. (2020). Blockchain technology, improvement suggestions, security challenges on smart grid and its application in healthcare for sustainable development. *Sustainable Cities and Society*, *55*, 102018.

[33] K. P. Satamraju and B. Malarkodi, "Proof of concept of scalable integration of internet of things and block chain in healthcare," *Sensors (Switzerland)*, vol. 20, no. 5, 2020, doi: 10.3390/s20051389.